\documentclass[prd,nofootinbib,superscriptaddress,twocolumn]{revtex4}

\usepackage{amsfonts,amssymb,amsthm,bbm}

\usepackage{amsmath}

\usepackage{hyperref}

\usepackage{color,psfrag}
\usepackage[dvips]{graphicx}
\usepackage{tikz}
\usetikzlibrary{calc}
\usetikzlibrary{decorations.pathmorphing}

\newcommand{\C}{{\mathbb C}}
\newcommand{\N}{{\mathbb N}}

\newcommand{\cB}{{\mathcal B}}

\newcommand{\cR}{{\mathcal R}}

\newcommand{\cV}{{\mathcal V}}
\newcommand{\cD}{{\mathcal D}}

\newcommand{\cI}{{\mathcal I}}

\newcommand{\SU}{\mathrm{SU}}

\newcommand{\SL}{\mathrm{SL}}
\newcommand{\SO}{\mathrm{SO}}

\newcommand{\be}{\begin{equation}}
\newcommand{\ee}{\end{equation}}
\newcommand{\beq}{\begin{eqnarray}}
\newcommand{\eeq}{\end{eqnarray}}
\newcommand{\bes}{\begin{eqnarray}}
\newcommand{\ees}{\end{eqnarray}}

\newcommand{\la}{\langle}
\newcommand{\ra}{\rangle}

\newcommand{\tr}{{\mathrm{Tr}}}
\newcommand{\f}{\frac}

\def\nn{\nonumber}

\def\tm{\tilde{m}}

\newcommand{\id}{\mathbb{I}}

\def\rd{\textrm{d}}

\def\centerarc[#1](#2)(#3:#4:#5)
    { \draw[#1] ($(#2)+({#5*cos(#3)},{#5*sin(#3)})$) arc (#3:#4:#5); }
    
\def\centerarcnodes[#1](#2)(#3:#4:#5)(#6,#7)
    {\coordinate(#6) at ($(#2)+({#5*cos(#3)},{#5*sin(#3)})$);
    \coordinate(#7) at ($(#2)+({#5*cos(#4)},{#5*sin(#4)})$);
    \draw[#1] ($(#2)+({#5*cos(#3)},{#5*sin(#3)})$) arc (#3:#4:#5); }

\def\angcircle(#1)(#2)(#3:#4) {\coordinate(#1) at ($(#2)+({#4*cos(#3)},{#4*sin(#3)})$); }

\begin{document}

\title{From Coarse-Graining to Holography in Loop Quantum Gravity}

\author{{\bf Etera R. Livine}}\email{etera.livine@ens-lyon.fr}
\affiliation{Univ Lyon, ENS de Lyon, Univ Claude Bernard, CNRS, LPENSL, 69007 Lyon, France}


\date{\today}

\begin{abstract}

We discuss the relation between coarse-graining and the holographic principle in the framework of loop quantum gravity and ask the following question: when we coarse-grain arbitrary spin network states of quantum geometry, are we integrating out physical degrees of freedom or gauge degrees of freedom?
Focusing on how bulk spin network states for bounded regions of space are projected onto boundary states, we show that all possible boundary states can be recovered from bulk spin networks with a single vertex in the bulk and a single internal loop attached to it.
This partial reconstruction of the bulk from the boundary leads us to the idea of realizing the Hamiltonian constraints at the quantum level as a  gauge equivalence reducing arbitrary spin network states to one-loop bulk states. This proposal of ``dynamics through coarse-graining'' would lead to a one-to-one map between equivalence classes of physical states under gauge transformations and boundary states, thus defining holographic dynamics for loop quantum gravity.


\end{abstract}

\maketitle


Loop quantum gravity sets up a non-perturbative framework for quantum gravity, with evolving quantum state of geometries and area and volume operators with quantized spectra at the Planck scale (for reviews of both the basic formalism and recents developments, see \cite{Thiemann:2006cf,Thiemann:2007zz,Rovelli:2011eq,Rovelli:2014ssa}). It faces a triptych of interlaced issues:
the coarse-graining of quantum geometry states from the Planck scale to larger scales, the definition of quantum dynamics consistent with the holographic principle and the implementation of (discretized) diffeomorphism  at quantum level as the fundamental gauge symmetry of the theory (or, in other words, the implementation of a relativity principle for quantum geometry).
These encompass more technical questions, such as anomaly cancellation, a well-behaved continuum limit and the perturbative renormalisation of quantum gravity corrections. In this short letter, we would like to discuss the relation between coarse-graining and holography.

First, there is the natural question of whether loop quantum gravity, and more generally any approach to quantum gravity, should be holographic.
On top on the area-entropy law for black holes, the related generalized entropy bounds for general relativity and the AdS/CFT correspondence at asymptotic infinity, the insight into holography is directly related to the invariance under diffeomorphism. This symmetry is at the heart of general relativity and is the mathematical translation of the relativity principle. However it is a tough challenge to identify and construct diffeomorphism-invariant observables (especially in pure gravity). Considering a bounded region allows to introduce a boundary, which acts on an anchor: looking for observables invariant under bulk diffeomorphisms leaving the boundary invariant seems to point towards the idea that all physical observables about the bulk geometry could/should be represented as boundary observables. We believe it is crucial, in order to understand better the structure and implications of loop quantum gravity, to investigate if the formalism can support such a (quasi-local) version of holography.

\smallskip

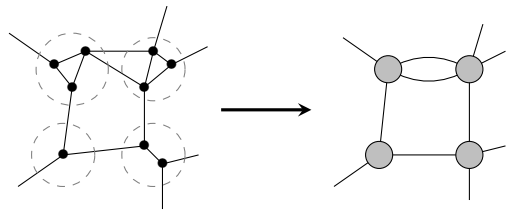
\begin{figure}[!h]
\centering

\begin{tikzpicture}[scale=0.6]

\coordinate(a1) at (-0.2,0);
\coordinate(a2) at (0.5,0.3);
\coordinate(a3) at (0.2,-0.5);
\coordinate(b1) at (2,0.3);
\coordinate(b2) at (1.8,-0.5);
\coordinate(b3) at (2.4,0);
\coordinate(c1) at (0,-2);
\coordinate(d1) at (1.8,-1.8);
\coordinate(d2) at (2.2,-2.2);

\draw (a1) -- (a2);
\draw (a1) -- (a3);
\draw (a3) -- (a2);
\draw (b1) -- (b2);
\draw (b1) -- (b3);
\draw (b3) -- (b2);
\draw (a2) -- (b1);
\draw (a2) -- (b2);
\draw (a3) -- (c1);
\draw (c1) -- (d1);
\draw (d1) -- (d2);
\draw (b2) -- (d1);
\draw (a1)--++(-1,.7);
\draw (c1)--++(-1,-.7);
\draw (b1)--++(.3,1);
\draw (b3)--++(.8,.4);
\draw (d2)--++(.8,.2);
\draw (d2)--++(0,-1);

\draw (a1) node {$\bullet$};
\draw (a2) node {$\bullet$};
\draw (a3) node {$\bullet$};
\draw (b1) node {$\bullet$};
\draw (b2) node {$\bullet$};
\draw (b3) node {$\bullet$};
\draw (c1) node {$\bullet$};
\draw (d1) node {$\bullet$};
\draw (d2) node {$\bullet$};

\draw[gray,dashed] (0.2,-0.1) circle(.8);
\draw[gray,dashed] (2,-0.1) circle(.7);
\draw[gray,dashed] (0,-2) circle(.7);
\draw[gray,dashed] (2,-2) circle(.7);

\draw[->,>=stealth,very thick] (3.5,-1) -- (5.5,-1);

\coordinate(a) at (7.2,-0.1);
\coordinate(b) at (9,-0.1);
\coordinate(c) at (7,-2);
\coordinate(d) at (9,-2);

\draw (a)--++(-1,.7);
\draw (c)--++(-1,-.7);
\draw (b)--++(.3,1);
\draw (b)--++(.8,.4);
\draw (d)--++(.8,.2);
\draw (d)--++(0,-1);

\draw (a) to[bend left] (b);
\draw (a) to[bend right] (b);
\draw (a) -- (c);
\draw (b) -- (d);
\draw (c) -- (d);

\draw[fill=lightgray] (a) circle(0.3);
\draw[fill=lightgray] (b) circle(0.3);
\draw[fill=lightgray] (c) circle(0.3);
\draw[fill=lightgray] (d) circle(0.3);

\end{tikzpicture}

\caption{A two-dimensional illustration of the partitioning of the graph in order to coarse-grain the spin network state.}
\label{fig:coarse-grain}
\end{figure}

Let us start by describing the standard setting for coarse-graining LQG. Quantum states of geometry are spin networks, that is graphs dressed with algebraic data: $\SU(2)$ representations, or spins, on the graph links and singlet states, or intertwiners, on the graph nodes. These spin networks define the bulk geometry, with the underlying graph representing a network of points as the backbone of the 3d space. A node represents an elementary chunk of volume and is usually thought of geometrically in terms of a dual surface surrounding it.
We now split up space into regions by partitioning the spin network into connected collections of nodes, as shown on fig.\ref{fig:coarse-grain}. The procedure is to coarse-grain each region to a single node, thus leading to a coarse-grained network describing the coarse-grained geometry of space.
The data attached to each coarse-grained node should reflect the information available on the geometry of the corresponding region of space:
we define a projector mapping quantum states of bulk geometry inside each region onto states living on the region's boundary surface.
This boundary surface  is thought of as the dual surface to the coarse-grained node and the projected boundary state is the new algebraic data attached to coarse-grained node.

Actually this is exactly the same framework as when discussing the implementation of the holographic principle in loop quantum gravity by describing the bulk geometry through the dynamics of holographic screens throughout space, except for an apparent essential discrepancy in the point of view:
coarse-graining means naturally loss of information, while the holographic principle claims that all the physically relevant degrees of freedom of the bulk are described by a surface theory without loss of information.
This is resolved by underlining a crucial difference in the premises: kinematics versus dynamics.

Indeed, the coarse-graining procedure is to be applied to arbitrary quantum spin network states, not necessarily satisfying the Hamiltonian constraints: the goal is to understand the large scale structure emerging from arbitrary Planck scale quantum geometries. On the other hand, the holographic principle is a statement about the quantum gravity dynamics and applies to physical states satisfying the Hamiltonian constraints. 
This leads to one big question: is the information, lost when coarse-graining by projecting the bulk geometry onto the boundary,  physically-relevant or is it pure gauge (for instance, under diffeomorphisms generated by the quantum Hamiltonian constraints)?
For example\footnotemark, in topological BF theory, all the detail of the graph within a region can be gauged out and physical spin network states exactly projected onto the topological defects living on its boundary.
\footnotetext{
Another example is provided by the distinction between bulk entropy and boundary entropy for black holes in loop quantum gravity \cite{Livine:2007sy}: holonomies wrapping around internal loops of the spin network graph within a region are non-trivial degrees of freedom generating entropy. If we want the bulk entropy to scale with the area at leading order, we should either bound the number of loops that can develop in the bulk or consider that these internal holonomies are partly gauge degrees of freedom. On the other hand, it was shown recently in \cite{Anza:2016fix,Anza:2017dkd} that, as soon as the number of internal loops is large enough, the boundary state is automatically thermal and respects the area-entropy law.
}
Actually, this natural interplay between holography and topological invariance was at the root of the proposal of considering general relativity as a constrained BF theory and realizing quantum gravity from topological quantum field theory (TQFT), which materialized into the spinfoam path integral for loop quantum gravity \cite{Barrett:1995mg,Barrett:1997gw,Markopoulou:1997hu,Barrett:2000xs}.

This logic  leads us to a drastic proposal to implement the dynamics of loop quantum gravity: we could consider all the information lost in the coarse-graining projection are gauge degrees of freedom and take this as a definition of the Hamiltonian constraints defining the diffeormophism invariance at the quantum level.
This can be understood as an extension of the topological invariance (in particular, of the triangulation invariance under Pachner moves) of BF theories to  theories, such as (quantum) gravity, which are non-topological but nevertheless holographic. 
This would be in spirit similar to the proposal of ``dynamics through coarse-graining'' by Dittrich and Steinhaus \cite{Dittrich:2013xwa} and sets coarse-graining procedures for spin network states as a fundamental building block of the theory.


Following this line of thought, we propose to analyze the projection from bulk geometry to boundary state in loop quantum gravity and focus on two points. First, can we classify\footnotemark{} all the bulk spin network states (graph and algebraic data) leading to a same boundary state? What is the ``redundant'' information stored in the bulk?
\footnotetext{
We could go further by formulating this question as a Kadison-Singer problem. Given a diagonal boundary state, attributing values to a maximal set of commuting observables probing the geometry of the boundary, what are its possible extension of a quantum state of the bulk geometry?
This could be especially interesting from the perspective of the relation between this question and the sparsification of networks, which is relevant to the issue of coarse-graining spin networks.
}
Second, what is the simplest bulk structure compatible with a given  boundary state?
We investigate these questions in the most straightforward LQG formulation, with quantum states of geometry defined as $\SU(2)$ spin networks and not as some notion of extended spin network\footnotemark{}.
\footnotetext{
Various extensions of spin networks have been introduced since the original formulation of loop quantum gravity in terms of $\SU(2)$ spin networks: $\SL(2,\C)$ simple and projected spin networks used in spinfoam path integrals \cite{Dupuis:2010jn,Rovelli:2010ed}, spin networks labeled with representations of the Drinfled double $\cD(\SU(2))$ to account for curvature and torsion excitations \cite{Dittrich:2016typ,Delcamp:2016yix}, double spin networks with holonomies both along the graph edge and looping around them \cite{Charles:2016xzi}  or the recently developed ``loop gravity string'' framework mixing the $\SU(2)$ algebraic structures with the Virasoro algebra of surface diffeomorphism living on each boundary surface \cite{Freidel:2015gpa,Freidel:2016bxd}.
}
We consider a bounded region of a spin network state, with the graph puncturing the boundary surface at $N$ points (see e.g. \cite{Feller:2017ejs} for a discussion on the definition of quantum surfaces in loop quantum gravity). Each puncture carries an arbitrary spin, so that the boundary state lives in the tensor product of $N$ arbitrary spins with the only constraint that the sum of all $N$ boundary spins is an integer:
\be
\cB_{N}
 =
\bigoplus_{\sum_{i=1}^Nj_{i}\in\N}\cB_{j_{1},..,j_{N}}
\subset
\left[\bigoplus_{j\in\f\N2}\cV_{j}\right]^{\otimes N}\,,
\ee
\be
\cB_{j_{1},..,j_{N}}
=
\bigotimes_{i=1}^{N} \cV_{j_{i}}\,,
\quad
\dim\cB_{j_{1},..,j_{N}}
=\prod_{i}^Nd_{j_{i}}\,,
\nn
\ee
where $j_{i}$ are the spins carried by the $N$ links puncturing the boundary surface and the dimension of the spin-$j$ representation $d_{j}=\dim\cV_{j}=2j+1$.
If the spin network graph in the bulk (here referring to the interior of the bounded region) is a tree, i.e. does not contain any loop, then it is known that the resulting boundary states are necessarily intertwiners, that is singlet states with vanishing overall spin. This clearly does not allow for arbitrary boundary state and can not be the generic case. So the question we will address in this letter is how much should we complicate the graph in order to allow for all possible boundary states.

\section{Boundary states for a single loop in the bulk}



Let us start by the simplest extension of a tree graph and consider a graph with a single loop.
We will show that we get all possible boundary states if we allow for an arbitrary $\SU(2)$ holonomy around the loop.

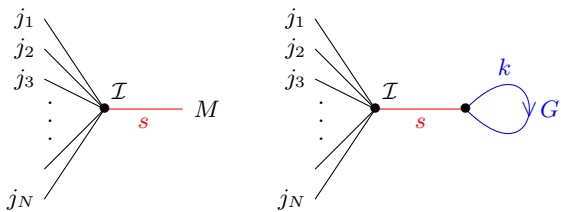
\begin{figure}[h!]

\centering

\begin{tikzpicture}[scale=0.8]

\coordinate(A) at (-1.5,0);
\coordinate(B) at (0,0);

\draw (A)--++(-1,1.5) node[left]{$j_{1}$};
\draw (A)--++(-1,1) node[left]{$j_{2}$};
\draw (A)--++(-1,0.5) node[left]{$j_{3}$};
\draw (A)--++(-1,-1);
\draw (A)--++(-1,-1.5) node[left]{$j_{N}$};

\draw (-2.4,0.1) node {.};
\draw (-2.4,-0.2) node {.};
\draw (-2.4,-0.5) node {.};

\draw[blue] (B) to[in=-45,out=+45,loop,scale=4] node[midway]{$\vee$} (B)++(1.4,0) node {$G$} (B)++(0.65,0.7) node {$k$};

\draw[red] (A) -- (B) node[midway,below]{$s$};
\draw (A) node {$\bullet$} ++(0.25,0.3) node {$\cI$};
\draw (B) node {$\bullet$};

\coordinate(C) at (-6,0);
\coordinate(D) at (-4.7,0);

\draw (C)--++(-1,1.5) node[left]{$j_{1}$};
\draw (C)--++(-1,1) node[left]{$j_{2}$};
\draw (C)--++(-1,0.5) node[left]{$j_{3}$};
\draw (C)--++(-1,-1);
\draw (C)--++(-1,-1.5) node[left]{$j_{N}$};

\draw (-6.9,0.1) node {.};
\draw (-6.9,-0.2) node {.};
\draw (-6.9,-0.5) node {.};

\draw[red] (C) -- (D) node[midway,below]{$s$};
\draw (C) node {$\bullet$} ++(0.25,0.3) node {$\cI$};
\draw (D)++(0.4,0) node{$M$} ;

\end{tikzpicture}

\caption{On the left hand side, we illustrate the basis states of the boundary Hilbert space $\cB_{N}$, as given in eqn.(\ref{Bs1}-\ref{Bs3}), labeled by the boundary spins $j_{1},..,j_{N}$, the closure defect spin $s$ and magnetic moment $M$ and intertwiner $\cI$ recoupling the boundary spins and the closure defect. On the right hand side, we draw spin network states for a one-loop bulk, as defined in eqn.\eqref{oneloop}, with the intermediate spin $s$ on the link between the boundary spins $j_{1},..,j_{N}$ and the internal loop carrying the spin $k$ and the holonomy $G\in\SU(2)$.} 
\label{fig:oneloop}

\end{figure}

A useful decomposition of the boundary Hilbert space is to partition it in terms of the closure defect \cite{Livine:2013gna,Anza:2016fix,Anza:2017dkd,Feller:2017ejs}. Mathematically, this means recoupling all the spins $j_{i}$ carried by the $N$ punctures into a single overall spin $s$. This spin, living on the intermediate channel, is necessarily an integer\footnotemark:
\be
\cB_{N}=\bigoplus_{s\in\N}\cB^{s}_{N}\,,
\quad
\cB^{s}_{N}=\bigoplus_{\{j_{i}\}_{i=1..N}}\cB^{s}_{j_{1},..,j_{N}}\,,
\label{Bs1}
\ee
\be
\cB^{s}_{j_{1},..,j_{N}}=
\cV_{s}\otimes \textrm{Inv}\Bigg{[}{\cV_{s}}\otimes\bigotimes_{i=1}^{N}\cV_{j_{i}}\Bigg{]}\,.
\label{Bs2}
\ee
\footnotetext{
A bounded region $\cR$ is a set of nodes of the spin networks, with all the links connected to them. We distinguish the interior links, whose both ends belong to $\cR$ and the boundary links with a single node in $\cR$. The parity condition at a node $v$ is that the sum of the spins around that node is an integer. Summing all the parity conditions for the vertices in $\cR$ implies that the sum of the boundary spins is necessarily an integer.
This holds as long as there is no source of torsion (e.g. fermions) within the region.
}
As depicted on fig.\ref{fig:oneloop}, a convenient basis of each subspace $\cB^{s}_{N}$ is given by choosing a basis state $|s,M\ra$ with magnetic moment $M$ in $\cV_{s}$ and an intertwiner $\cI$ recoupling the $N$ spins $j_{i}$ with the  spin $s$:
\be
\cI\in\textrm{Inv}\Bigg{[}{\cV_{s}}\otimes\bigotimes_{i=1}^{N}\cV_{j_{i}}\Bigg{]}\,,\,\,
\la s,M|\cI\ra\in\cB^{s}_{j_{1},..,j_{N}}
\,.
\label{Bs3}
\ee

Now let us look at spin network states with boundary spins $j_{i}$ based on the single node graph with a self-loop,  as drawn on fig.\ref{fig:oneloop}. A convenient basis is to separate the loop from the boundary edges, choose a spin $k$ carried by the loop and once more intertwine the boundary spins $j_{i}$ into a single overall spin $s$. Then the spin network states is defined by two intertwiners, one defining the recoupling of the $j_{i}$'s to $s$ and one defining the recoupling of two copies of the spin $k$ into $s$. The latter is actually unique once $k$ and $s$ are given since it is a 3-valent node.
The boundary states defined by these spin network basis states   are expressed in terms of the holonomy $G\in\SU(2)$ carried by the loop and the intertwiner $\cI\in\cB^{s}_{j_{1},..,j_{N}}$ :
\be
\psi_{\{j_{i}\}}^{s,k,\cI}[G]
=
D^k_{m,\tm}(G)
C^{k,s|k}_{\tm,M|m}
\la s,M|\cI\ra
\,\in
\cB_{j_{1},..,j_{N}}\,.
\label{oneloop}
\ee
with the Wigner-matrix of the holonomy,
\be
D^k_{m,\tm}(G)=
\la k,\tm|G|k,m\ra
\,,
\nn
\ee
and the Clebsh-Gordan coefficient recoupling three spins:
\be
C^{k,s|k}_{\tm,M|m}
=
\la k,m|(k,\tm)\otimes(s,M)\ra
\,.
\nn
\ee
We would like to prove that these states $\psi_{\{j_{i}\}}^{s,\cI}[G]$ cover the whole boundary space $\cB_{j_{1},..,j_{N}}$ for any given boundary spins $j_{1},..,j_{N}$. We compute\footnotemark{} the density matrix $\rho_{k}$ integrating over the holonomy $G$ carried by the loop at fixed loop spin $k$:
\beq
\rho_{k}&=&\int \rd G\,|\psi_{\{j_{i}\}}^{s,k,\cI}[G]\ra\la\psi_{\{j_{i}\}}^{s,k,\cI}[G]|
\\
&=&
\int \rd h\,
{\chi_{k}(h)^2}
D^s_{MM'}(h)
\la s,M|\cI\ra\la \cI | s,M'\ra \nn\,,
\eeq
%
\footnotetext{
We use the useful realization of the product of two Clebsh-Gordan coefficients as an integral over the product of three Wigner-matrices:
\be
C^{k,s|k}_{\tm,M|m}
\overline{C^{k|k,s}_{m'|\tm',M'}}
=
(2k+1)
\int\rd h\,
D^s_{MM'}(h)
D^k_{\tm\tm'}(h)
D^k_{m'm}(h^{-1})
\,.
\nn
\ee
}
where $\chi_{k}(h)=\tr\,D^{k}(h)$ is the character of the group element $h\in\SU(2)$ for the spin-$k$ representation.
An interesting formula\footnotemark{} for distributions over $\SU(2)$ is:
\be
\delta_{\SO(3)}(h)
=
\sum_{n\in\N}(2n+1)\chi_{n}(h)
=
-2\sum_{k\in\f\N2}\chi_{k}^2(h)\,.
\ee
\footnotetext{
We distinguish the $\delta$-distribution on $\SU(2)$, which expands over all spins in $\N/2$, and the $\delta$-distribution on $\SO(3)$, which involves only integer spins. Then we use the tensor product formula for characters, recoupling two copies of a spin $k$:
\be
\forall k\in\f\N2\,,\,\,\chi_{k}^{2}=\sum_{n=0}^{2k}\chi_{n}\,,
\quad
\forall n\in\N^{*}\,,\,\,\chi_{n}=\chi_{\f n2}^{2}-\chi_{\f{n-1}2}^{2}\,.
\nn
\ee
}
Since the spin $s$ in the intermediate channel is always an integer, we can sum over the loop spin $k$ and obtain the density matrix $\rho$:
\be
\rho
=-\f12\sum_{k}\rho_{k}
=\la s,M|\cI\ra\la \cI | s,M\ra
=\id^{s}_{j_{1},..,j_{N}}\,,
\ee
which we recognize as the identity of the boundary Hilbert space $\cB^{s}_{j_{1},..,j_{N}}$.

This not only shows that the boundary states induced by a one-loop bulk cover the whole boundary Hilbert space, but it also provides an explicit decomposition of the identity on the boundary space in terms of intertwiners with one self-loop. This allows to decompose an arbitrary boundary state in $\cB_{N}$ in terms of those one-loop spin networks, thus providing them with an interpretation of coherent bulk states.

\section{From boundary states to bulk states and vice-versa}

The mathematical result proved above has a direct application to the question of the local reconstruction of the bulk geometry from the boundary state in LQG.
The minimal reconstruction does not require a complex graph structure in the bulk, we only need a single loop in the bulk.
Using the formula for the decomposition of the identity on $\cB^{s}_{j_{1},..,j_{N}}$ in terms of one-loop bulk states, an arbitrary boundary state unfolds on a graph with one internal loop and determines the (probability amplitude for the) values of the internal holonomy $G$, internal spin $k$ living on the loop and the closure defect spin $s$.
From this point of view, the boundary state does not carry further information on the bulk structure and it seems that we should consider any more complicated graph structure in the bulk as gauge degrees of freedom.

This can be put in contrast with the loopy spin network framework developed in \cite{Charles:2016xwc} in an attempt to define a consistent coarse-graining procedure for loop quantum gravity, proposing to consider the dynamics of spin networks on a fixed graph background but allowing for an arbitrary number of self-loops attached to each node. These self-loops (or little loops as named in \cite{Charles:2016xwc}) are interpreted as representing local excitations of curvature (localized at each node of the background graph).
Here, our result seems to collapse this multi-loop structure and require a single loop at each graph node to deal with the coarse-graining of the theory.

So we formulate a very crude proposal, which will necessarily need to be refined:
the Hamiltonian constraints imposing diffeomorphism invariance can be implemented in loop quantum gravity as a gauge invariance rendering all bulk graph and spin network states to be gauge-equivalent to a one-loop bulk state. 
Although this would drastically simplify the theory, it does not reduce to a topological BF theory and is clearly an extension beyond it:
in BF theory, all internal loops are pure gauge and can be entirely gauged-out, while here we allow for one-loop states  and thus for local curvature and local degrees of freedom living on those loops.
This is consistent with the formulation of general relativity as a BF theory with extra constraints, which is the standard starting point for constructing spinfoam path integrals for implementing the dynamics of loop quantum gravity \cite{Barrett:1997gw,DePietri:1998hnx,Reisenberger:1998fk,Engle:2007wy} (see \cite{Perez:2012db} for a review).

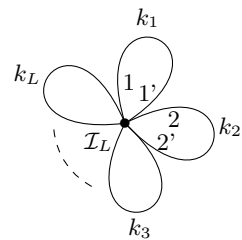
\begin{figure}[h!]
\centering
\begin{tikzpicture}
\coordinate(O1) at (0,0);


\draw (O1) to[in=-37,out=+37,loop,scale=4,rotate=70] (O1) ++(0.3,1.4) node {$k_1$};
\draw (O1) to[in=-37,out=+37,loop,scale=4,rotate=-10] (O1) ++(1.4,-0.05) node {$k_2$};
\draw (O1) to[in=-37,out=+37,loop,scale=4,rotate=-80] (O1) ++(0.2,-1.4) node {$k_3$};
\draw (O1) to[in=-37,out=+37,loop,scale=4,rotate=150] (O1) ++(-1.3,0.65) node {$k_{L}$};

\draw (O1) node[scale=1] {$\bullet$} ++(-0.35,-0.25) node{$\cI_{L}$};

\draw (O1) ++(0.05,0.55) node{1};
\draw (O1) ++(0.3,0.4) node{1'};
\draw (O1) ++(0.65,0.05) node{2};
\draw (O1) ++(0.55,-0.25) node{2'};

\centerarc[dashed](O1)(185:245:.95);

\end{tikzpicture}
\caption{A spin network state on a flower graph, with a single central vertex and $L$ loops/petals: the $L$ spins $k_{1}$,..,$k_{L}$ live on pair of half-edges and are recoupled by the intertwiner $\cI_{L}$.}
\label{fig:flower}
\end{figure}

Let us try to be more explicit and use the simplified setting of flower graphs to illustrate our approach.  We consider graph with a single vertex and $L$ loops attached to it, as on fig.\ref{fig:flower}, as introduced to the context of loop quantum gravity in \cite{Freidel:2002xb,Charles:2016xwc}. We could attach exterior legs carrying spins $j_{i}$ but they will not be coupled to the spins on the loops in our examples of dynamics, although there will clearly be an possibility to investigate further in future work. The spin network states on a flower graph with $L$ loops or petals are labelled by spins $k_{1}$,.., $k_{L}$ and a central intertwiner $\cI_{L}$ living in their tensor products:
\be
|\{k_{1},..,k_{L}\},\cI_{L}\ra\,,\quad
\cI_{L}\in \textrm{Inv}_{\SU(2)}\Big{[}
\bigotimes_{l=1}^{L}
\cV^{k_{l}}\otimes \overline{\cV^{k_{l}}}
\Big{]}\,.
\ee
Let us call $l,l'$ the two half-edges of each loop connected to the vertex.
The BF dynamics is encoded in the equivalence relation between spin networks:
\be
|\{k_{1},..,k_{L}\},\cI_{L}\ra
\,\underset{BF}{\sim}\,
|\{k_{1},..,k_{i}-\f12,..,k_{L}\},\widetilde{\cI_{L}}\ra
\,,
\ee
where the new intertwiner $\widetilde{\cI_{L}}$ is obtained from the initial one $\cI_{L}$ through a $F_{ll'}$-operator, as defined in the $U(N)$ framework for $\SU(2)$-intertwiners \cite{Freidel:2010tt,Borja:2010rc}, which lowers the spins $k_{i}$ by $\f12$ while preserving the $\SU(2)$-invariance. Let us choose $i=L$. Then, as $k_{L}$ decreases, we eventually get $k_{L}=0$, which is equivalent to no loop $L$ at all:
\be
|\{k_{1},..,k_{L}\},\cI_{L}\ra
\,\underset{BF}{\sim}\,
|\{k_{1},..,k_{L-1}\},\widetilde{\cI_{L-1}}\ra\,,
\ee
where the reduced intertwiner $\widetilde{\cI_{L-1}}$ is obtained from $\cI_{L}$ by contracting it with the states on the two half-edges $L,L'$ glued by the identity group element $\id$.
This equivalence can be realized through the action of (modified) holonomy operators, as shown in detail in \cite{Charles:2016xwc}, and ultimately means that all loops can be removed: all flower states are equivalent and they are equivalent to the trivial vaccuum state with no loop at all.

We would like to put this topological case in contrast with a non-trivial example where the coarse-grained state retain information on the bulk geometry. We introduce an ``spin-preserving'' , or ``area-preserving'', equivalence relation, which couples pair of loops:
\be
|\{..,k_{i},..,k_{j},..\},\cI_{L}\ra
\,\underset{new}{\sim}\,
|\{..,k_{i}+\f12,..,k_{j}-\f12,..\},\widetilde{\cI_{L}}\ra
\,,
\nn
\ee
where the new intertwiner $\widetilde{\cI_{L}}$ is obtained from $\cI_{L}$ by a product operator $F_{ii'}^{\dagger}F_{jj'}$, increasing the spin $k_{i}$ and  lowering the spin $k_{j}$ while preserving the $\SU(2)$-invariance. This equivalence relation preserving the total area around the vertex, $\sum_{l}k_{l}$. Ultimately, setting as many spins as possible to 0, we are left with a single loop carrying the total area:
\be
|\{k_{1},..,k_{L}\},\cI_{L}\ra
\,\underset{new}{\sim}\,
|\{\sum_{l=1}^{L}k_{l}\}\ra
\,,
\ee
where the bivalent intertwiner, for the one-loop  case, is unique. This provides a first example in the simplified setting on flower graphs of non-trivial loop quantum gravity dynamics, such that spin network states are all dynamically-equivalent to single loop bulk states.

\medskip

In order to make this proposal explicit in general, we need the operator(s) at the quantum level which generate(s) the gauge invariance under diffeomorphism between arbitrary graphs and states and map(s) any bulk spin network state onto the corresponding one-loop bulk state. Here we follow the coarse-graining procedure, ``coarse-graining by gauge-fixing'' defined in detail in \cite{Livine:2006xk,Livine:2013gna,Charles:2016xwc} to project any bulk state onto the corresponding  boundary state and then use the formula presented in the previous section to map that boundary state onto the corresponding one-loop bulk state. More generally, we can generalize this logic to any coarse-graining procedure defined for kinematical spin network states and turn it into a definition of the gauge invariance implementing the dynamics of spin networks in loop quantum gravity. This underlines the fundamental physical relevance of a coarse-graining procedure for spin networks. And it underlines the need for a systematic study of coarse-graining procedures in order to classify them and analyze the universality of their flows.

Finally, this proposal would automatically implement the holographic principle in loop quantum gravity: physical states for the bulk geometry would be exactly gauge-equivalent to one-loop bulk states and thus, in other words, gauge-equivalent to boundary states.

\section*{Outlook \& Conclusion}

We proved that it is possible to interpret all boundary states, on a quantum surface with $N$ punctures,  as spin network states on a bulk graph consisting in a unique vertex dressed with an intertwiner and a single little loop attached to it and carrying an arbitrary $\SU(2)$ holonomy.
This led us to the idea of ``dynamics by coarse-graining'' for loop quantum gravity, where the Hamiltonian constraints would translate to gauge transformations mapping arbitrary bulk spin network states to one-loop bulk states. This would mean that physical states (solving the Hamiltonian constraints) would be in a one-to-one correspondence with boundary states. It would be a straightforward extension of the dynamics of BF theory: (little) loops would not be completely pure gauge but almost, so that we gauged-out all internal loops but one for any bounded region of space.
Although this is, for now, more the outline of a proposal  than an explicit realization, this framework would automatically implement the holographic principle in the heart of loop quantum gravity.

This proposal implicitly means that there is no diffeomorphism-invariant observable - that is an operator commuting with the quantum Hamiltonian constraints - measuring the geometry of a bounded finite region of space that is sensitive to fine details to the bulk graph structure. Identifying such an observable (e.g. such that its spectrum measures the number of bulk vertices or bulk loops) would be a great progress for loop quantum gravity and would imply that the Hilbert space of physical states inside a bounded region be larger than just one-loop bulk states. This could be investigated in recent proposals of Hamiltonian constraints for loop quantum gravity \cite{Alesci:2011ia,Assanioussi:2015gka}.

To go further along this line of investigation, the next step would be to check whether our conclusion is robust to the proposed extensions of spin network states, for instance if we consider both curvature and torsion defects with spin networks labeled with Drinfeld double representations (based on exponentiated fluxes) as in \cite{Dittrich:2016typ,Dittrich:2017nmq} or if we add the information of the Virasoro currents living on the surfaces (encoding information on  the intrinsic geometry of the surface) as advocated in \cite{Freidel:2016bxd}.
Considering the latter proposal, it will be very interesting to understand how the coarse-graining procedure for spin networks could lead to conformal field theory states on the boundary and how such a proposal for the loop quantum gravity dynamics fits with the  possibility of a local gravity/CFT correspondence.
%

Finally, it would be essential to understand the relation between the point of view we developed and the recent result by Anz\`a and Chirco \cite{Anza:2017dkd} that a complicated-enough bulk (defined by a graph with a large enough number of internal loops) would typically lead to thermal boundary states. This points towards the necessity of a finer analysis of the bulk-to-boundary relation in loop quantum gravity. For instance, we can formulate the issue of  {\it purification} in quantum geometry: considering an arbitrary {\it mixed boundary state} defined by an arbitrary density matrix, can we identify it as the projection of a {\it pure bulk state} defined on a potentially complicated bulk graph?

%

\vfill

%
%


\bibliographystyle{bib-style}
\bibliography{LQG}

\end{document}